\documentclass[11pt]{article}

\usepackage{amsmath}

\newcommand{\N}{N\raise.7ex\hbox{\underline{$\circ $}}$\;$}

\begin{document}

\title{ E.M. Ovsiyuk\footnote{e.ovsiyuk@mail.ru}\\
 Particle in the  Electromagnetic Wave with Cylindrical Symmetry,
 an Analog of the Volkov Problem
 \\
{\small Mozyr State Pedagogical University named after I.P. Shamyakin, Belarus}}

\date{}

\maketitle

\begin{abstract}

Cylindrically symmetric analogue of the Volkov problem is
examined. In presence of the field  of a  cylindrical
electromagnetic wave,  classical motion of a non-relativistic
particle on a cylindrical surface   is described exactly in terms
of elliptic functions.

\end{abstract}

\section*{1. Particle in a plane wave, the Volkov  problem  }

Problems of mathematical physics constitute a basis for simulating
processes in modern physics.
 Expanding the range of such problems is useful in many ways.
  In particular, in the context of investigating the properties of nanostructures
  may be of interest the  transport of electrical charges in a very narrow cylindrical conductive
   layer in  presence of an external electromagnetic  wave with cylindrical symmetry.
   As shown in this study,  such a system has many common mathematical properties
   of   the well-known  Volkov  problem   [1]  on the motion of electric charge in the ordinary
   plane electromagnetic wave.

Let us start with the potential for a plane wave propagating along
positive direction of the $z$-axis
$$
A_{1}(x^{3}) = A \cos {2\pi \over T} (t - {x^{3} \over c} ) = A
\cos (\omega t - k  x^{3})\; ,
$$
$$
A_{0}=0\;, \;
 A_{2} = 0\; , \;  A_{3} = 0 \; , \; k = {\omega \over c} \; ;
$$
$$
 F_{01} = -A k  \; \sin (\omega t - k  x^{3})\; ,
\qquad
F_{31} = + A k \; \sin (\omega t - k  x^{3}) \; . \eqno(1.1)
$$

Transforming  this field to the  cylindrical coordinates
$$
x^{1} = r \cos \phi \; , \qquad x^{2} = r \sin \phi\;, \qquad
x^{3}
 = z\; ,
 $$
 $$
 dS^{2}= c^{2} dt^{2} -  d r^{2} -  r^{2} \; d \phi^{2}  -
dz^{2}\;  ,
\eqno(1.2a)
$$

\noindent
we get
$$
A_{r} = { \partial x^{1} \over \partial r} A_{1} = +A   \; \cos
\phi  \cos (\omega t - kz)\; , \qquad A_{0}= 0\; , \qquad
$$
$$
A_{\phi} = { \partial x^{1} \over \partial \phi} A_{1} = -A  \; r
\sin \phi  \cos (\omega t - kz) \; , \qquad A_{z} = 0 \; ;
\eqno(1.2b)
$$
$$
F_{0r} = - A k \cos \phi   \sin (\omega t - kz)\; ,
\;
F_{0\phi} = + A  k \; r \cos \phi   \sin (\omega t - kz)\; ,
\;F_{0z} = 0 \; ,
$$
$$
F_{\phi z} =  k  A \;  r  \sin \phi  \sin (\omega t - kz) \; ,
\;
F_{zr} = k   A   \; \cos \phi  \sin (\omega t - kz) \; , \;
F_{r\phi} = 0 \; .
$$
$$
\eqno(1.2c)
$$

Next we use the  shortening notation for the phase:
$
\omega t - k z = \Omega$;
 then a plane wave is described by the relations
$$
A_{r} = +A   \; \cos \phi  \; \cos \Omega \; , \qquad A_{\phi} =
-A  \; r  \sin \phi  \; \cos \Omega \; ,
$$
$$
F_{0r} = - A k \cos \phi   \; \sin \Omega\; , \qquad F_{0\phi} = +
A  k \; r \sin \phi   \; \sin \Omega \; ,
$$
$$
F_{\phi z} =  k  A \;  r  \sin \phi  \; \sin \Omega \; , \qquad
F_{zr} = k   A   \; \cos \phi  \; \sin \Omega  \; . \eqno(1.3)
$$

Let us consider the problem of a particle in the plane wave, using
Lagrangian formalism (from the very beginning specifying the problem in cylindrical coordinates)
 [2]
$$
L = {m \over 2} \;   (- g_{ik} V^{i}V^{k} )  -  {e  \over c}  \;
g_{ik} A^{i}V^{k}  =
$$
$$
=
{m \over 2} \; ( \;   V^{r}  V^{r}  + r^{2} \; V^{\phi}
V^{\phi}
 +   V^{z} V^{z} \; ) -
 $$
 $$
 - {e \over c }   A   \; \cos \phi  \; \cos \Omega  \;  V^{r}  +
 {e \over c}  A  \; r  \sin \phi  \; \cos \Omega \;  V^{\phi} \; .
\eqno(1.4)
$$

\noindent Euler--Lagrange equations take the form
$$
{d \over  dt} \; (\;    V^{r}   - {eA \over mc}       \; \cos \phi
\; \cos \Omega  \;   ) =
  r V^{\phi}\ V^{\phi} +
 {eA \over mc}   \;  \sin \phi   \cos \Omega   V^{\phi}\; ,
 $$
$$
 {d \over dt} \; ( \ r^{2} V^{\phi} +  {eA \over mc}     r  \sin \phi   \cos \Omega \; )=
 {eA \over mc}       \sin \phi   \cos \Omega    V^{r}  +
{eA \over mc}    \; r  \cos \phi   \cos \Omega \;  V^{\phi} \; ,
$$
$$
 {d \over dt}   V^{z} =
 - {eA \over mc}     k \; \cos \phi  \; \sin \Omega  \;  V^{r}  +  {eA \over mc}
  k \; r  \sin \phi  \; \sin \Omega \;  V^{\phi}\; .
$$
$$
 \eqno(1.5)
$$

\noindent
The system (1.5) is equivalent to
$$
  {d V^{r}  \over d t   }  - r V^{\phi} V^{\phi}   = - {eA  \over m}   k \; \cos \phi \; \sin \Omega  +
 {eA \over m c}  k \cos \phi\;  \sin \Omega
 \; V^{z}      \; ,
$$
$$
{d  V^{\phi}  \over d t  } +
 {2 \over r} \; V^{r} V^{\phi}  =  {eA \over m}  k  \; {  \sin \phi \over r}  \; \sin \Omega  -
 {eA \over mc}  k  \; { \sin \phi \over r} \;  \sin \Omega  \;  V^{z}   \; ,
 $$
 $$
{d  V^{z}  \over d t  }   =  - {eA \over mc}  k  \;\cos \phi  \;
\sin \Omega \;  V^{r}  +
 {eA \over m c} k  \; r \sin \phi \; \sin \Omega \; V^{\phi}      \; .
\eqno(1.6)
$$

A simpler system of equations can be obtained by starting with
Cartesian coordinates
$$
m {d V^{1} \over  dt } = e E^{1} + e ( V^{2} B^{3} - V^{3} B^{2})
\; ,
$$
$$
m {d V^{2} \over  dt } = e E^{2} + e ( V^{3} B^{1} - V^{1} B^{3})
\; ,
$$
$$
m {d V^{3} \over  dt } = e E^{3} + e ( V^{1} B^{2} - V^{2} B^{1})
\; . \eqno(1.7)
$$

\noindent  Taking into account
$$
 E^{1} = F_{01} = -A k \; \sin (\omega t - k  x^{3})  \; ,
 $$
 $$
 c B^{2} = -F_{31} = - A k \; \sin (\omega t - k  x^{3}) \; ,
$$

\noindent equations (1.7) are written
$$
{d V^{1} \over  dt } = - {eA\over m}  k  \; \sin \Omega   +    {eA
\over mc} k \;    V^{3} \; \sin \Omega \; ,
$$
$$
{d V^{2} \over  dt } =  0  \; , \qquad {d V^{3} \over  dt } =   -
{eA  \over mc} k \;  V^{1}  \;   \sin \Omega  \; . \eqno(1.8)
$$

Systems (1.8) and (1.6) must be equivalent, which is readily verified  by direct recalculation,
  using the formulas
$$
V^{1} = -r \sin \phi \; V^{\phi}  + \cos
\phi \; V^{r}\; , \qquad
V^{2} =  r \cos \phi \; V^{\phi} + \sin
\phi \; V^{r} \; .
\eqno(1.9)
$$

 Next we will analyze the equations (1.8). According to
(1.8), along the $ x ^ {2} $ the particle moves with constant velocity.
In the plane of the $ 1 - $ 3, its motion is described by equations
$$
{1 \over c } {d V^{1} \over  dt } = - {eA\over mc}  k  \; \sin
\Omega   +    {eA  \over mc} k \;    {V^{3} \over c}  \; \sin
\Omega \; ,
$$
$$
{1 \over c} {d V^{3} \over  dt } =   - {eA  \over mc} k \;   \sin
\Omega  \; {V^{1}  \over c}  \;    . \eqno(1.10)
$$

\noindent
Let us introduce the variables
$$
q = {eA\over mc}  k \; , \qquad  [\; q \; ] = {1 \over sec}\; ,
\qquad v^{1} = {V^{1} \over c}\; , \qquad  v^{2} = {V^{2} \over c}
\eqno(1.11)
$$

\noindent then equations (1.10) can be written as
$$
{d v^{1} \over  dt } = - q  \; \sin \Omega   +   q \;    v^{3}
\; \sin \Omega \; , \qquad
 {d v^{3} \over  dt } =   - q \;     \sin \Omega \; v^{1} \; .
\eqno(1.12)
$$

You can find an approximate solution of this equation,   considering the motion for sufficiently small
time intervals  (see [3]):
$$
v^{3}   << 1 \; , \qquad
 k z = {\omega \over c} \int _{0}^{t} {d z \over dt} \; dt = \omega \; \int {V^{3} \over c} \; dt << \omega t
\eqno(1.13)
$$

\noindent  that is, by imposing an additional constraint
$
\Omega = \omega t - k z \approx \omega t \; .
$
This system of equations (1.12) takes the form
$$
{d v^{1} \over  dt } = - q  \; \sin \omega   t \; , \qquad
 {d v^{3} \over  dt } =   - q \;     \sin \omega t \; v^{1} \; .
\eqno(1.14)
$$

\noindent
Integrating the first equation, we get
$$
v^{1} = {q  \over \omega} \; \cos \omega t + (v^{1}_{(0)}  - {q
\over \omega} ) \; ,
$$
$$
 {x \over c}  = {q \over \omega^{2}}
\sin \omega t  +  (v^{1}_{(0)} - {q \over \omega})  \; t + {x_{0}
\over  c} \; . \eqno(1.15)
$$

\noindent After that, it is easy to integrate the second equation
$$
{d v^{3} \over  dt } =   - q \;  [  \; {q \over \omega}  \; \cos
\omega t + (v^{1}_{(0)}-{q \over \omega}  ) \; ] \;   \sin \omega
t  \; ,
$$
$$
v^{3} = - {q^{2} \over 2  \omega^{2} } \sin^{2} \omega t  +
{q\over \omega  } (v^{1}_{(0)}- {q \over \omega})  \; \cos \omega
t +  [\; v^{3}_{(0)} - {q\over  \omega } (v^{1}_{(0)}- {q \over
\omega}) \; ] \; ,
$$
$$
{z \over c}   = - {q^{2} \over 2 \omega^{2} } ( { t \over  2} - {
\sin 2\omega t \over 4 \omega})   + {q\over \omega ^{2} }
(v^{1}_{(0)}- {q \over \omega})
 \; \sin \omega t +
 [\;
v^{3}_{(0)} - {q\over \omega } (v^{1}_{(0)}- {q \over \omega} ) \;
] \; t + {z_{0} \over c}  \; .
$$
$$
\eqno(1.16)
$$

Let us turn to the system  (1.12) in the general case,
and translate it  to a new time variable $\tau$
$$
\tau = t - {z \over c} \; , \qquad t = \tau + {z \over c} \; ,
\qquad dt = d \tau + {dz \over c}\; . \eqno(1.17a)
$$

\noindent
Generalized velocities transform according to
$$
v^{1} = { c^{-1} d x^{1}   \over d \tau  + c^{-1} dz }= {
\hat{v}^{1} \over 1 +   \hat{v}^{3}} \; , \qquad v^{2} = { c^{-1}
d x^{2}  \over d \tau  + c^{-1} dz }= { \hat{v}^{2} \over 1 +
\hat{v}^{3}} \; ,
$$
$$
\hat{v}^{1} = c^{-1} {d x^{1} \over d \tau}\; , \qquad \hat{v}^{2}
= c^{-1} {d x^{2} \over d \tau}\; , \qquad \hat{v}^{3} =  c^{-1}
{d x^{3} \over d \tau}\; ,
$$
$$
v^{3} = { c^{-1} d z  \over d \tau  + c^{-1} dz }= {\hat{v}^{3}
\over 1 +   \hat{v}^{3}} \; , \qquad \hat{v}^{3} =  { v^{3} \over
1 -  v^{3}} \; ; \eqno(1.17b)
$$

\noindent operator of differentiation with respect to time is converted
according to
$$
{d \over dt}  f = {d \over d \tau + c^{-1} d z} f = {1 \over 1 +
\hat{v}^{3}} \; {d \over d \tau} f \; . \eqno(1.17c)
$$

\noindent
The system of equations (1.12) takes the form
$$
{1 \over 1 +   \hat{v}^{3}} \; {d \over d \tau} {\hat{v}^{1} \over
1 +   \hat{v}^{3}}
  = -q    \; \sin \omega \tau   +    q \; \sin \omega \tau \;
   {\hat{v}^{3} \over 1 +   \hat{v}^{3}} \; ,
$$
$$
{1 \over 1 +   \hat{v}^{3}} \; {d \over d \tau} {\hat{v}^{3} \over
1 +   \hat{v}^{3}}  =   -  q \; \sin \omega \tau\; {\hat{v}^{1}
\over 1 +    \hat{v}^{3}}  \;     \;  .
\eqno(1.18)
$$

\noindent
The previous equation can be rewritten as
$$
{1 \over q  \sin   \omega \tau}\;  {1 \over 1 +   \hat{v}^{3}} \;
{d \over d \tau} \; ( { -\hat{v}^{1} \over 1 +   \hat{v}^{3}} )
  =  1      -       {\hat{v}^{3} \over 1 +   \hat{v}^{3}}  = {1 \over 1 + \hat{v}^{3}} \; ,
$$

$$
{1 \over  q \sin \omega \tau\;}\; {1 \over 1 +  \hat{v}^{3}} \; {d
\over d \tau} {\hat{v}^{3} \over 1 +   \hat{v}^{3}}  =  ( -
{\hat{v}^{1} \over 1 +  \hat{v}^{3}}  )  \;     \;  . \eqno(1.20)
$$

\noindent
Allowing for the  second equation, from the first one we get
$$
\left ( {1 \over  q \sin   \omega \tau}\;  {1 \over 1 +
\hat{v}^{3}} \; {d \over d \tau} \; \right )  \left (\;
 {1 \over  q \sin \omega \tau\;}\; {1 \over 1 +  \hat{v}^{3}} \; {d \over d \tau}\; \right )\;
{\hat{v}^{3} \over 1 +   \hat{v}^{3}}
  = {1  \over 1 +   \hat{v}^{3}}    \;
\eqno(1.21a)
$$

\noindent or

$$
{d \over d \tau} \; \left [\;
 {1 \over  \sin \omega \tau\;}\; {1 \over 1 +  \hat{v}^{3}} \; {d \over d \tau}
{\hat{v}^{3} \over 1 +   \hat{v}^{3}}  \; \right ]
  =   q ^{2} \; \sin \omega \tau  \; .
\eqno(1.21b)
$$

\noindent
After integrating over $\tau $ we have
$$
 {1 \over 1 +  \hat{v}^{3}} \; {d \over d \tau}
{\hat{v}^{3} \over 1 +   \hat{v}^{3}}
  = (  -{ q^{2}  \over \omega} \; \cos \omega \tau   + \lambda  )\; \sin \omega \tau   .
\eqno(1.21c)
$$

\noindent
Note that in the limit
$
\hat{v}^{3}  << 1 \; , \;  k z << \omega t \; , \; \tau
\approx t \; , \;  \hat{v}^{3} \approx v^{3},$ the previous equation is simplified
$$
  {d \over d t}
v^{3}
  = (  -{ q ^{2}  \over \omega} \; \cos \omega t   + \lambda  )\; \sin \omega t \;  ,
\eqno(1.22)
$$

\noindent which coincides with the first equation in (1.16).

We return to equation ($1.21c$):
$$
 {1 \over 1 +  \hat{v}^{3}} \; {d \over d \tau}
{\hat{v}^{3} \over 1 +   \hat{v}^{3}}
  = (  -{ q^{2}  \over \omega} \; \cos \omega \tau   + \lambda  )\; \sin \omega \tau   .
\eqno(1.23a)
$$

\noindent
It is a differential equation of first order
with separated  variables
$$
{d \; \hat{v}^{3} \over (1 + \hat{v }^{3} )^{2}} =  (  -{ q^{2}
\over \omega} \; \cos \omega \tau   + \lambda  )\; \sin \omega
\tau   \; d \tau \; , \eqno(1.23b)
$$

\noindent which is easily integrated (we introduce a constant
integration)
$$
- {1 \over 2( 1 + \hat{v}^{3})^{2} } = - {q^{2} \over \omega}  \;
{\sin^{2} \omega \tau \over  2 \omega} - {\lambda \over  \omega}
\;  \cos \omega \tau - { \Lambda  \over 2 \omega^{2}}
\;
 \eqno(1.23c)
$$

\noindent and further
$$
\hat{v}^{3} =  -1 \pm  { \omega  \over \sqrt{
 q^{2}  \; \sin^{2} \omega \tau
+ 2\lambda  \omega   \;  \cos \omega \tau +  \Lambda   } } \; .
\eqno(1.23d)
$$

From ($1.23d$)  one can obtain an expression for the particle velocity
$v^{3}$ -- see ($1.17b$):
$$
{1 \over c}\; {d z \over dt} = v^{3}  = { \hat{v}^{3} \over  1 +
\hat{v}^{3}} \; ,
$$
$$
{d z \over dt}  = c \mp\; {c \over  \omega}\; \sqrt{
 q^{2}  \; \sin^{2} \omega (t - z/c)
+ 2\lambda  \omega   \;  \cos \omega (t - z/c) +  \Lambda   }  \; \; .
\eqno(1.24a)
$$

If we assume that at the time $ t = 0 $ coordinate $ z = z_ {0} = 0 $,
then from ($1.24a$) it follows
$$
\dot{z}_{0}= c \mp\; {c \over  \omega}\; \sqrt{
 2\lambda  \omega    +  \Lambda   } \; .
\eqno(1.24b)
$$

Variant with the lower sign must be discarded as not physical,
because it involves an initial velocity greater than the rate of
of light.
We return to equation ($1.23d$)
$$
c^{-1} {dz \over d \tau} =   -1 \pm  { \omega  \over \sqrt{
 q^{2}  \; \sin^{2} \omega \tau
+ 2\lambda  \omega   \;  \cos \omega \tau +  \Lambda   } } \; ,
$$

\noindent so we get

$$
u = \omega \tau = \omega( t - z/c) \; ,
$$

$$
{z \over c}  = - \tau \pm \int \; {  d u \over \sqrt{
 q^{2}  + \Lambda    - q^{2} \cos^{2} u  + 2\lambda  \omega   \;  \cos u     } }  \; .
\eqno(1.25)
 $$

This integral reduces to an elliptic one (in more detail  we consider the analogous integral in the next section).
   As a result we get the function of $ z (t) $ in an implicit transcendental form --
this is a known property of the classical problem of motion of a charged particle in a plane electromagnetic wave
(Volkov [1]).

In the next section we will construct a cylindrically symmetric analog of such a  system,
treating  a particle in the simplest plane electromagnetic wave with cylindrical symmetry.

\section*{ 2. Particle in the presence of a  cylindrically \\ symmetric electromagnetic  wave}

We start from a simple solution of the Maxwell's equations in cylindrical
coordinates  $(ct, r, \phi, z)$
$$
A_{0}=0, \qquad A_{r} = 0\;, \qquad A_{\phi} = a\;  \cos (\omega t
- kz) \; , \qquad A_{z} = 0 \; ; \eqno(2.1)
$$

\noindent it  meets the following electromagnetic tensor
$$
F_{0 \phi} = -  k  a \; \sin (\omega t - kz)\; , \qquad F_{z \phi}
= + k  a \; \sin (\omega t - kz)\; . \qquad \eqno(2.2)
$$

\noindent It is easy to check that here indeed  we have  the solution of the
Maxwell's equations
$$
{1 \over \sqrt{-g} } {\partial \over \partial x^{\alpha} }
\sqrt{-g} F^{\alpha \beta} = 0\; ,
$$

\noindent which is reduced to one non-trivial
equation
$$
\beta = \phi  \; , \qquad   - c \partial_{t} {1 \over r^{2}}
F_{0 \phi}
 +    \partial_{ z } {1 \over r^{2}}   F_{z  \phi}  = 0 \;  \qquad  \Longrightarrow
$$
$$
- c^{-1}  \partial_{t}   [ -  k  a \;  \sin (\omega t - kz)    ]
 +    \partial_{ z }   [ k  a \; \sin (\omega t - kz) ]   = 0 \; ,
 $$
\noindent that is an identity
$$
+ k^{2} a\; \cos (\omega t - kz) -  k^{2} a\; \cos (\omega t - kz)
=  0 \; .
$$

It is useful to convert the electromagnetic 4-potential to Cartesian
coordinates
$$
A_{1} = - { x^{2}
\over  (x^{1})^{2} +
 (x^{2})^{2} } \; a  \; \cos (\omega t - k z) \; ,
 $$
 $$
 A_{2} =  +{ x^{1}
\over  (x^{1})^{2} +  (x^{2})^{2} } \; a \; \cos (\omega t - kz)
\; . \eqno(2.3)
$$

\noindent It corresponds to the electromagnetic tensor
$$
cB^{1} =  F_{23} =    - { x^{1} \over  (x^{1})^{2} +  (x^{2})^{2}
} \; k a  \;  \; \sin (\omega t - k z)\; ,
 $$
 $$
 cB^{2} = F_{31} =   - { x^{2} \over  (x^{1})^{2} +
 (x^{2})^{2} } \; k a  \;  \; \sin (\omega t - k z)\; ,
$$
$$
E^{1} = F_{01} =  { x^{2} \over  (x^{1})^{2} +
 (x^{2})^{2} } \; ka  \; \sin (\omega t - k z)\;,
 $$
 $$
E^{2} = F_{02} =  - { x^{1} \over  (x^{1})^{2} +
 (x^{2})^{2} } \; ka  \; \sin (\omega t - k z)\;,
$$
$$
\qquad cB^{3} = F_{12} = 0\; , \qquad  \qquad E^{3}= F_{03}=0 \; .
 \eqno(2.4)
 $$

\noindent This is a wave propagating along the axis $ z $, it is singular
on the axis $ z $ (when  $ r = 0 $). In the plane of the fixed value of the variable $ z $,
at each point the magnetic field is directed toward the center $ (0,0) $ along
the radius,  and the electric field at each point in the plane is
directed along the vector $ {\bf e} _ {\phi} $. Amplitude of the electric and magnetic fields
varies according to the same law
$$
E  \sim {1 \over r} \;ak\;  \sin (\omega t - k z)\;  , \qquad   c
B \sim  {1 \over r} \; ak\;  \sin (\omega t - k z) \; .
$$

Consider a particle in that field.  The Lagrangian of the system is given by
 (examining the problem of a non-relativistic particle)
$$
L = {m \over 2} ( \; -g_{ij}(x)  V^{i} V^{j}  ) - {e \over c}  \;
g_{ij}(x) \; A^{i} V^{j}=
$$
$$
= {m \over 2 }\;  [ V^{r} V^{r} + r^{2} V^{\phi} V^{\phi}  +
V^{z} V^{z} ) -{e \over c}  \; a\; \cos (\omega t - kz) V^{\phi}
\; . \eqno(2.5)
$$

\noindent Euler--Lagrange equations
$$
{d \over dt} {\partial L \over \partial  V^{r}}  = {\partial L
\over \partial  r }\; , \qquad {d \over dt} {\partial L \over
\partial  V^{\phi}}  = {\partial L \over \partial  \phi }\; ,
\qquad {d \over dt} {\partial L \over \partial  V^{z}}  =
{\partial L \over \partial  z }\; ,
$$

\noindent here take the form
$$
{d V^{r} \over dt} = r \; V^{\phi} V^{\phi} \; ,
$$
$$
{d \over dt}\;  [ \; r^{2} V^{\phi} - {ea \over  m c}  \; \cos
(\omega t - kz) \; ] = 0 \qquad  \Longrightarrow \qquad   {d I \over dt } = 0\; ,
$$
$$
{d V^{z} \over dt } = - { ea \over m c } k \; \sin  (\omega t -
kz) \;  V^{\phi}\; . \eqno(2.6)
$$

From (2.6) we get a  much more simpler systems, if
 consider the motion at sufficiently small time intervals:
$$
 k z = {\omega \over c} \int _{0}^{t} {d z \over dt} \; dt = \omega \; \int {V^{3} \over c} \; dt << \omega t
\eqno(2.7a)
$$

\noindent  that is, by imposing an additional constraint
$$
\omega t - k z \approx \omega t \; . \eqno(2.7b)
$$

\noindent Then (2.6) gives (we use the notation $ea /mc
= b $)
$$
{d V^{r} \over dt} = r \; V^{\phi} V^{\phi} \; , \qquad
 V^{\phi}    = {I  + b  \; \cos \omega t\; \over r^{2}} \;  ,
 $$
 $$
 {d V^{z} \over dt } = - b k \; \sin  \omega t  \;  V^{\phi}\; .
\eqno(2.8)
$$

\noindent and then we get three equations with separated
variables:
$$
{d ^{2} r \over dt^{2}} =   { (I  + b  \; \cos \omega t )^{2}
\over r^{3}}   \; ,
$$
$$
{d V^{z} \over dt } = - b k \;   { \sin  \omega t  \; (I  + b  \;
\cos \omega t ) \over r^{2}(t)} \; ,
$$
$$
 V^{\phi}    = {I  + b  \; \cos \omega t\; \over r^{2}(t)} \;  .
\eqno(2.9)
$$

Unfortunately, such a system hardly can be solved as well.
One can  try to simplify the original system by imposing additional const\-raint.
The most interesting from a physical point of view, is the condition $r=r_{0} =
\mbox{const}$.
It means that a particle can only move  along a given cylindrical surface.
 In this case the electric component of the external wave speed varies
 linearly along the vector ${\bf e}_{\phi}$,
  and the magnetic field changes the linear velocity along the axis $z$.

With this  condition, instead of (2.6) we  will have a simpler Lagrangian and two new equations of motion:
$$
L = {m \over 2 }\;  [r^{2}_{0} V^{\phi} V^{\phi}  + V^{z} V^{z} )
-{e \over c}  \; a\; \cos (\omega t - kz) V^{\phi} \; ,
$$
$$
{d \over dt}\;  [ \; r^{2}_{0} V^{\phi} -  b  \; \cos (\omega t -
kz) \; ] = 0 \; ,
$$
$$
{d V^{z} \over dt } = - b  k \; \sin  (\omega t - kz) \;
V^{\phi}\; .
\eqno(2.10)
$$

\noindent
Variable $ \phi $ varies with time according to
$$
V^{\phi} =  { I + b \; \cos (\omega t - kz)  \over r_{0}^{2}} \; .
\eqno(2.11a)
$$

\noindent The second equation in (2.10) takes the form
differential equation for  $ z (t) $

$$
{d V^{z} \over dt } = - {b  k \over r_{0}^{2}} \;     \sin
(\omega t - kz) \; [\; I + b \; \cos (\omega t - kz) \; ]
  \; .
\eqno(2.11b)
$$

First we will find the solution of equations (2.11) in the approximation of the velocity
much smaller than the speed of light, and not too large
time intervals

$$
 k z = {\omega \over c} \int _{0}^{t} {d z \over dt} \; dt = \omega \; \int {V^{3} \over c} \; dt << \omega t\; ,
\qquad \omega t - k z \approx \omega t \; .
$$

\noindent In this case (2.11) can be simplified
$$
V^{\phi} =  { I + b \; \cos \omega t   \over r_{0}^{2}} \; ,
\qquad V^{\phi}_{0} = {I  + b \over r_{0}^{2}} \; ,
$$
$$
{d V^{z} \over dt } = - { b  k \over r_{0}^{2}}  \; \sin  \omega t
\;   ( I + b \; \cos \omega t   ) \; . \eqno(2.12)
$$

\noindent
These equations are easily integrated:

$$
\phi (t) =   \phi_{0} + {I \over r_{0}^{2} }  \; t +  {b \over
r_{0}^{2} \omega}  \; \sin \omega t \,, \eqno(2.13a)
$$

$$
V^{z}  = - { b  k \over r_{0}^{2}}  \; [\; {b \over 2 \omega} \;
\sin^{2}  \omega t  \;   - {I \over \omega} \; \cos \omega t   \;
] + ( V^{z}_{0} -  \; { b  k  I \over r_{0}^{2} \omega } ) \;  ,
$$

$$
z(t)  = \int \{ - { b  k \over r_{0}^{2}}  \; [\; {b \over 2
\omega} \;  { 1 - \cos 2  \omega t \over 2}   \;   - {I \over
\omega} \; \cos \omega t   \; ] + ( V^{z}_{0} -  \; { b  k  I
\over r_{0}^{2} \omega } ) \;  \} \; dt =
$$
$$
= - { b  k \over r_{0}^{2}}  \; [\; {b \over 2 \omega} \;
({1\over 2} t  - {\sin 2  \omega t \over 4 \omega}  )  \;   -
 {I \over \omega^{2}} \; \sin \omega t   \; ] +
( V^{z}_{0} -  \; { b  k  I \over r_{0}^{2} \omega } ) \; t +
z_{0} \; .
$$
$$
\eqno(2.13b)
$$

Character of the motion is as follows:
along the $ z $ axis  the particle moves with a certain average
constant speed, with two superimposed oscillatory
movement in time with frequencies $ \omega $ and $ 2 \omega $; on
variable $ \phi $ is also the movement at a constant
(angular) velocity and when imposed  oscillating motion.

Let us return to the analysis of equation ($2.11b$) without approximation of small
intervals of time. It is convenient to introduce  a new time variable
$$
\tau = t - {z \over c} \; , \qquad t = \tau + {z \over c} \; ,
\qquad dt = d \tau + {dz \over c}\; . \eqno(2.14a)
$$

\noindent
Generalized velocity $ V ^ {z} $ is transformed according to
$$
V^{z} = {  d z  \over d \tau  + c^{-1} dz }= {\hat{V}^{z} \over 1
+   c^{-1}  \hat{V}^{z}} \; , \qquad \hat{V}^{z} =  { V^{z} \over
1 -   c^{-1}  V^{z}} \; ; \eqno(2.14b)
$$

\noindent operator of differentiation with respect to time is converted
according to
$$
{d \over dt}  f = {d \over d \tau + c^{-1} d z} f = {1 \over 1 +
c^{-1} \hat{V}^{z}} \; {d \over d \tau} f \; . \eqno(2.14c)
$$

\noindent Equation ($2.11b$) takes the form
$$
{1 \over 1 +  c^{-1} \hat{V}^{z}} \; {d \over d \tau} \; { c^{-1}
\hat{V}^{z} \over 1 +   c^{-1}  \hat{V}^{z}}= - {b  k  \over  c }
\;    {  \sin  \omega \tau   (I + b \; \cos \omega \tau )  \over
r_{0}^{2}} \; ,
$$

\noindent or (go to the dimensionless velocity $\hat{v}^{z} =
c^{-1 } \hat{V}^{z}$)
$$
d \; \left ( {     1  \over   (1 +   \hat{v}^{z})^{2} }  \right )
=
 { 2  b  \omega  \over  c^{2} r_{0}^{2}} \;
   \sin  \omega \tau  \;  ( I + b \; \cos \omega \tau )   \;   d\tau \; .
\eqno(2.15a)
$$

In fact, this equation (up to notation)
coincides with ($1.23b$) -- an equation arising in the case of particles in the field of
an ordinary  plane wave. After
integration of ($2.15a$)
$$
{     1  \over   (1 +   \hat{v}^{z})^{2} }= { 2  b  \omega  \over
c^{2} r_{0}^{2}} \;   [ \;  {  b\over  2\omega }  \;  \sin^{2}
\omega \tau   -
 {I \over \omega} \;
   \cos  \omega \tau  \;  ] + C
$$

\noindent
and further
$$
{     1  \over   1 +   \hat{v}^{z}  }= \pm \sqrt{ { 2  b  \omega
\over  c^{2} r_{0}^{2}} \;   [ \;  {  b\over  2\omega }  \;
\sin^{2}  \omega \tau   -
 {I \over \omega} \;
   \cos  \omega \tau  \;  ] + C  } = \pm \sqrt{\Gamma (\tau) } \; .
\eqno(2.15b)
$$

The equation for the velocity $ \hat {v}^{z} $ can be integrated:

$$
c^{-1} \; {d z \over d \tau}  = \hat{v}^{z} =  -1 \pm {1 \over
\sqrt{\Gamma( \tau) }}  \qquad \Longrightarrow
$$
$$
z (\tau ) = - c \tau \pm \int {d \tau \over  \sqrt{\Gamma( \tau)
}}  \; .
\eqno(2.15c)
$$

There will arise an implicitly defined  function $ z (t) $ with
use of elliptic functions. Indeed, consider the
integral
$$
J = \int { d w \over \sqrt{a\cos^{2} w + b \cos w +c}} \; ;
\eqno(2.16a)
$$

\noindent introducing  a variable
$$
-\cos w = u\; ,  \; \;  dw = {d u
\over \pm \sqrt{1- u^{2}}} \; ,
$$
we get
$$
J = \int { du \over \sqrt{(1 - u^{2})(a u^{2} + b u +c )}}\,.
\eqno(2.16b)
$$

It is convenient to introduce yet another  variable  trough a linear fractional transformation

$$
u =  { \rho \; x + \sigma  \over x +1} \; ,
$$
$$
1 - u^{2} = { (1- \rho^{2})x^{2} + 2 (1 - \rho \sigma) x + (1 - \sigma ^{2}) \over
(1 + x)^{2}} \; ,
\eqno(2.17a)
$$
$$
a u^{2} + b u +c  = { (a \rho^{2} + b \rho + c) x^{2}   + [2a \rho \sigma + b (\rho + \sigma)
+2c ] x + (a \sigma^{2} + b \sigma + c)
 \over  (1 + x)^{2} } \; .
$$

\noindent
The required coefficients are determined by conditions
$$
1 - \rho \sigma = 0 \; , \qquad 2a \; \rho \sigma + b \; (\rho + \sigma)
+2c= 0 \; ,
$$
$$
\rho =   {  -(a+c) +  \sqrt{ (a+c)^{2} - b^{2} }   \over b } \; ,
$$
$$
\sigma =   {  -(a+c) -  \sqrt{ (a+c)^{2} - b^{2} }   \over b } \,.
\eqno(2.17b)
$$

\noindent
As a result, the integral $ J $ is the following:
$$
J = {1 \over A}\; \int { d x \over  \sqrt{1 - \rho^{2} x^{2}} \; \sqrt{ 1 + m^{2} x^{2} }}\; ,
$$
$$
A = { a + \rho b + \rho ^{2} c \over  \rho^{2} (\rho^{2} -1)} \; , \qquad
m^{2} = { \rho^{2} ( \rho^{2} \; a + \rho b + c    ) \over  a + \rho  b + \rho^{2} c   } \; .
\eqno(2.18)
$$

\noindent
Making the change of variables
$$
\rho \; x = \sqrt{1 - U^{2}}\; , \qquad k = { m \over \sqrt{\rho^{2} + m^{2}}} \; ,
$$

\noindent we reduce the integral to the canonical form of the elliptic integral:
$$
J = -{ 1 \over  \sqrt{A \; ( \rho^{2} + m^{2} )}} \; \int { d U \over \sqrt{1-U^{2}}\; \sqrt{1 - k^{2} U^{2}}}
\;.
\eqno(2.19)
$$

Therefore, as a result, we produce solution $ z (t) $ in an implicit form

$$
z (\tau ) = - c \tau   \pm     {1 \over \omega } \left [
\; -{ 1 \over  \sqrt{A \; ( \rho^{2} + m^{2} )}} \; \int { d U \over \sqrt{1- U^{2}}\; \sqrt{1 - k^{2} U^{2}}}
\; \right ]\,.
\eqno(2.20)
$$

\vspace{5mm}

Author is grateful to V.M. Redkov and Yu.A. Kurochkin  for advice and for help in writing this paper.

\vspace{10mm}

\end{document}